\documentstyle[prl,aps,epsfig,twocolumn]{revtex}

\begin{document}
  \columnsep0.1truecm
 %\draft
 %\preprint{ }

  \title  {Quantum Particle-Trajectories and Geometric Phase}
  \author {M. Dima}
  \address{Dept. of Physics, Campus Box 390            \\
           University of Colorado,                     \\
           Boulder, CO-80309                                }
  \maketitle

  \begin{abstract}
 ``Particle"-trajectories are 
 defined as integrable $dx_\mu dp^\mu = 0$
 paths in projective space.
 Quantum states
 evolving on such trajectories, open or closed,
 do not delocalise in $(x, p)$ projection,
  the phase associated with the trajectories being
 related to the
 geometric (Berry) phase and the 
 Classical Mechanics action.
 Properties at high energies of the states evolving
 on ``particle"-trajectories 
 are discussed.
\end{abstract}


\narrowtext

  Quantal wave-packet revival~\cite{ref:revival-th} 
 is the periodic re-assembly of a state's localised structure
  along a classically stable orbit. The phenomenon has
 been observed experimentally 
 in Rydberg
 atoms~\cite{ref:rydberg} as well as in one-atom 
 masers~\cite{ref:masers}, and prompts
 the question whether such revival
 is possible also for states evolving on open trajectories~\cite{ref:xpati},
 similarly to
 classical point-particles. 
  It is shown in this Letter that integrable
 $dx_\mu dp^\mu = 0$ 
 trajectories in projective space
 do provide such a context, the 
   aspect being related to the Differential
 Geometry properties of manifolds~\cite{ref:kobayashi},
 independent of 
 the existence
 of a Hamiltonian.
 
    The revival of
 quantal wave-packets is connected
 to the concept of
 {\em geometric phase}~\cite{ref:reviews} introduced by
 Berry. Berry~\cite{ref:Berry} has shown
 that additionally to a Hamiltonian induced
 dynamic phase, a quantum state evolving
 in parameter space on a trajectory that returns to the initial 
 state acquires an extra phase termed   geometric 
 phase. Subsequent analysis has generalised the context
 in which the phenomenon occurs, lifting
 the restriction of adiabaticity~\cite{ref:non-adiabat},
 cyclicity and unitarity~\cite{ref:non-unitar}.
 An important step was made by the 
 kinematic approach~\cite{ref:kinem},
 which demonstrated that the Hamiltonian is not needed
 in defining the geometric phase, and underlined
 the native geometrical nature of the quantity by
 relating it to the Bargman invariants~\cite{ref:bargman,ref:bargman2}.
  The acquirment of a geometrical phase
 by quantum states evolving on closed trajectories in 
 parameter space has
 been  
 verified experimentally in neutron
  interference~\cite{ref:neutron},  
 in two photon states produced in spontaneous
 parametric down-conversion~\cite{ref:laser}, {\em etc}. The latter
 paper~\cite{ref:laser} makes also the important remark
 that
 experiments related to non-locality {\em vis \`{a} vis}
   the Bell inequalities~\cite{ref:Bell} 
 and the Berry phase are connected, non-locality
 in Quantum Mechanics being pointed
 out as a consequence
 of completeness as 
 early as 1948 by
 Einstein~\cite{ref:einstein}.

 The sole assumptions of this Letter are that
 quantum systems are described by a linear representation
 space over ${\mathbf C}$~\cite{ref:dima}
 and that the coordinate operator ${\mathbf x}_\mu$
 has a {\em conjugate operator},
 $[{\mathbf x}^\mu, 
 {\mathbf k}^\nu]_{_{-}} =
 -ig^{\mu \nu} \cdot {\mathbf 1}$.
 The latter operators act as tangent space
 vectors on the manifold, action  revealed by
 the (Weyl) translation operators
 ${\mathbf U}_{\Delta x}  \, \buildrel \rm def \over = \,  
   e^{+i \Delta x_\mu
 {\mathbf k}^\mu}$ and 
 ${\mathbf U}_{\Delta k}  \, \buildrel \rm def \over = \, 
   e^{-i \Delta k_\mu
 {\mathbf x}^\mu}$ :
 \begin{eqnarray}
   {\mathbf U}^\dagger_{\Delta x} \, {\mathbf x}^\mu \, 
 {\mathbf U}_{\Delta x}
 &=& {\mathbf x}^\mu + \Delta x^\mu  \cr \cr
  {\mathbf U}^\dagger_{\Delta k} \, {\mathbf k}^\mu \,
 {\mathbf U}_{\Delta k}
 &=& {\mathbf k}^\mu + \Delta k^\mu
 \end{eqnarray} respectively
  $| x \rangle  
 \buildrel \rm {{\mathbf U}_{\Delta x}} \over \rightharpoondown
 | x + \Delta x \rangle$ and $| k \rangle  
 \buildrel \rm {{\mathbf U}_{\Delta k}} \over \rightharpoondown
 | k + \Delta k \rangle$.
 Given an arbitrary reference state $| \psi_{ref} \rangle$,
 a set of translated image-states
 can be defined as~\cite{ref:coherent}:
 \begin{equation}
 | \psi (\xi, \kappa) \, \rangle 
  \, \buildrel \rm def \over = \, 
 {\mathbf U}_{\Delta k} {\mathbf U}_{\Delta x} | \psi_{ref}
 \rangle
 \end{equation}
  with correspondingly translated state averages:
 \begin{eqnarray}
 \langle {\mathbf x}^\mu \rangle_{\psi(\xi, \kappa)}  &=&
 \langle {\mathbf x}^\mu
 \rangle_{ref} + \Delta x^\mu = \xi^\mu \cr \cr
 \langle {\mathbf k}^\mu \rangle_{\psi(\xi, \kappa)} &=& \langle
 {\mathbf k}^\mu \rangle_{ref} + \Delta k^\mu = \kappa^\mu
 \end{eqnarray}
 The spread of the image states is
 identical to that of the reference state,
 regardless
 the $(\Delta x, \Delta k)$ translation:
 \begin{eqnarray}
 \delta x^\mu_{\psi(\xi, \kappa)}  &=&  \, \, \,
 \delta x^\mu_{ref} = const. \cr \cr
   \delta k^\mu_{\psi(\xi, \kappa)} &=& \, \, \,
 \delta k^\mu_{ref} = const.
 \end{eqnarray}
  The interchange of 
  ${\mathbf U}_{\Delta x}$ and ${\mathbf U}_{\Delta k}$ 
    in the definition of the image state
 $| \psi (\xi, \kappa) \, \rangle$ 
 leads to a state corresponding in
 projective space~\cite{ref:non-adiabat,ref:bargman2} to the same point,
 the difference between the two being
 just a phase factor:
 \begin{equation}
 {\mathbf U}_{\Delta x} {\mathbf U}_{\Delta k} =
  e^{+i \Delta x_\mu \Delta k^\mu}
  {\mathbf U}_{\Delta k} {\mathbf U}_{\Delta x}
 \label{eq:openx}
 \end{equation}
 The situation
 is better evidentiated by the comparison
 of $| \psi_{ref} \rangle$
 with its transported
 image around a $\Delta x$ $\rightarrow$ 
 $\Delta k$ $\rightarrow$ $-\Delta x$ $\rightarrow$ $-\Delta k$
 quantum loop:
 \begin{equation}
 {\mathbf U}_{loop} = {\mathbf U}^\dagger_{\Delta k}
  {\mathbf U}^\dagger_{\Delta x} {\mathbf U}_{\Delta k} 
 {\mathbf U}_{\Delta x}  = 
  e^{-i \Delta x_\mu \Delta k^\mu}
 \cdot  {\mathbf 1}
 \label{eq:loopx-kl}
 \end{equation}
  respectively around an arbitrary quantum loop:
  \begin{eqnarray}
  {\mathbf U}_{loop}  =
  \prod_{loop} {\mathbf U}_{dk} {\mathbf U}_{dx} \, &=& \,
  e^{-i\oint k_\mu dx^\mu}
 \cdot {\mathbf 1} \cr \cr
  &=& \,
  e^{+i\oint x_\mu dk^\mu}
 \cdot {\mathbf 1}
 \label{eq:loopx} 
 \end{eqnarray} 
 In both cases the state acquires a geometrical
 phase proportional to the $(x, k)$ area enclosed
 by the loop in projective space.
  Should this phase be
 zero, the anholonomy~\cite{ref:non-unitar} hold
 preventing the realisation of
 a proper $(x, k)$ coordinate system
 on the representation space disappears, as it will be
 shown in the next paragraph.
 Generalising equation   (\ref{eq:openx}) to
 continuous open paths (for k$_i$ = 0):
   \begin{equation}
 {\mathbf U}_{open} =
  \prod_{initial}^{final}
 {\mathbf U}_{dk} {\mathbf U}_{dx} \, = \,
  e^{-i\int_i^f k_\mu dx^\mu}
 \cdot  {\mathbf U}_{\Delta k} 
 {\mathbf U}_{\Delta x}
   \label{eq:openy} 
  \end{equation}     
 and holding the initial and final
 states apart at fixed displacements $(\Delta x, \Delta k)$,
 a path dependent
 geometric phase for {\em open paths} can be defined,
 arbitrary up to a path independent gauge~\cite{ref:gauge}
   field $\Phi(x, k)$:
 \begin{equation}
  S  \, \buildrel \rm def \over = \,
 - \int_i^f k_\mu dx^\mu = + \int_i^f x_\mu dk^\mu
 \end{equation}
  The above relation 
 supports a class of canonical transformations
 (such as $Q = k$, $K = -x$) consistent
 with 
  $[{\mathbf x}^\mu, {\mathbf k}^\nu]_{_{-}} =
 -ig^{\mu \nu} \cdot {\mathbf 1}$ 
 and 
 $\langle x | k \rangle = 
 (2\pi)^{-2} e^{-i x_\mu k^\mu}$,
 that identifies geometrical
 phase as the Classical 
 Mechanics {\em action}~\cite{ref:arnold}.
  Assuming that $| \psi_{ref} \rangle$
 can evolve on two neighbouring paths via a beam-splitter 
 like mechanism, the interference
 in the final state is destructive unless $\delta S = 0$
 (for remote trajectories: $\delta S = 2n\pi$),
 respectively the {\em extremal
 action condition}.
  Paths 
 satisfying the extremal action condition
 at each point -
 or equivalently, in equation (\ref{eq:loopx-kl}) $dx_\mu dk^\mu = 0$ -
 preserve constructive interference along the path, and are
 termed ``{\em particle-trajectories}".
  This is not an exclusive
 category however, non-particle {\em infodynamics} being
 equally possible~\cite{ref:st}.
 The early attempts to formulate Quantum Mechanics
 in terms of $(x, p)$ coordinates failed due to the non-zero
 commutator of the coordinate and momentum operators
  $[{\mathbf x}^\mu, {\mathbf p}^\nu]_{_{-}} =
 -i\hbar g^{\mu \nu}  \cdot {\mathbf 1}$, and are
 best summarised by the Heisenberg inequality 
 $\delta x^\mu \cdot \delta p^\nu \ge \frac{\hbar}{2}|g^{\mu \nu}|$.
 Nonetheless, free propagation of quantum systems can be approximated
 by Classical Mechanics, 
 as hinted by the extremal geometric phase relation above.
 
  Establishing an $(x, k)$
 coordinate system on a manifold requires that
 a translation with a $\Delta x$ leg followed by one
 with a $\Delta k$ leg reach the same point
 as it would under those operations
 interchanged:
 \begin{equation}
   [{\mathbf U}_{\Delta x}, {\mathbf U}_{\Delta k}]_{_{-}} =
 (1 - e^{-i \Delta x_\mu \Delta k^\mu}) 
 {\mathbf U}_{\Delta x} {\mathbf U}_{\Delta k} = 0
 \end{equation}
 This is possible non-trivially only for spaces at least
 2D in
 dimension, by requiring $\Delta x_\mu \Delta k^\mu = 0$.
 The problem of establishing an $(x, k)$ grid
 on a 1D manifold is that a translation around
 a quantum loop of area $dx \cdot dk = \frac{1}{2}$
  accumulates a phase factor $\pi$,
 as seen from equation (\ref{eq:loopx}).
 For manifolds of greater dimension
 this phase may
 vanish by {\em reciprocal phase compensation}
 among dimensions of opposite metric sign. For an Euclidian metric
 it can be shown that the condition is met
 only by trajectories on the sphere, while for
 the Minkowski metric non-trivial
 solutions of the $n + 1$ pairs of canonically
 conjugate variables - $(Q, K)$
 plus the {\em temporal} dimension $(T, H)$ - are allowed.
 To have thus a proper $(x,k)$ coordinate system on
 the manifold two conditions must be met:
\begin{enumerate}
\item{ - {\bf necessary condition}: $dx_\mu dk^\mu =
  0_{\big \vert_{PATH}}$

 This relation defines {\em locally} a coordinate system, and
 it is better known in physics than apparent at first glance.
 For example in the case of wave-packet propagation,
 requiring the constituent waves to move in sync
 yields
 the condition $\vec{v}_g = \vec{\nabla}_{\! k} \omega$,
 which re-written as
 $\vec{v}_g \cdot d\vec{k} =
  \vec{\nabla}_{\! k} \omega \cdot d\vec{k} = d\omega$, 
 becomes:
      \begin{equation}
      dt \cdot d\omega - d\vec{x} \cdot d\vec{k} = 0
     \end{equation}
 For  point-particles, the  work-energy
 relation $dE = \vec{F} d\vec{x} = d\vec{x} \cdot d\vec{p}/dt$
   can likewise be re-written as:
    \begin{equation}
      dt \cdot dE - d\vec{x} \cdot d\vec{p} = 0
     \end{equation}
 }
\item{ - {\bf sufficient condition}: $ d^2x = 0$ and 
                                     $d^2k = 0_{\big \vert_{PATH}}$
     
  This relation conditions path
 integrability, necessary for the
 path independent definition of an $(x, k)$ coordinate system on the
 manifold.
 It is a {\em global} condition,
 the standard
 solution~\cite{ref:demonstratie} being - up to a canonical 
 transformation~\cite{ref:arnold}:
 \begin{eqnarray}
  k_\mu k^\mu &=& {\pm k_C^2} \cr \cr
  \frac{d x^\mu}{\| dx \|}  &=&  \frac{k^\mu}{k_C}
 \end{eqnarray}
 The traditional ``dynamical" character of ${\mathbf k}_\mu$
 stems precisely from this solution,
 and less so from its more distantly related
 Differential Geometry properties on the manifold.
  The inertia of the differential equations rules out
 ``cross-over"
 trajectories from $k_\mu k^\mu > 0$ to $k_\mu k^\mu < 0$ paths,
 $\pm k_C^2$
 being a characteristic of the trajectory.
 Likewise, trajectories
 on the light-cone cannot
 ``fall" onto $k_\mu k^\mu > 0$ or $k_\mu k^\mu < 0$
 solutions either, due to the gradient of
 the differential equation parallel to the sheet
 of the light-cone.   The $k_\mu k^\mu = \pm k_C^2$
 relation is also well 
  known in physics, in the form of $E = c 
 \sqrt{m_0^2c^2+\vec{p}^{\, 2}}$,
  respectively:
  \begin{equation}
   (E/c)^2 - \vec{p}^{\, 2} = (m_0c)^2
  \end{equation}
   }
 \end{enumerate}
   In summary, up to a canonical
 transformation~\cite{ref:arnold}
 ``particle"-trajectories provide a ruling of the manifold
 that satisfies the:
\begin{itemize}
\item{ translational
 properties of state averages:
\begin{eqnarray}
   \langle {\mathbf x}^\mu \rangle_{\psi(\xi, \kappa)}
  &=& \langle {\mathbf x}^\mu
  \rangle_{ref} +
                    \Delta x^\mu \cr \cr
  \langle {\mathbf k}^\mu \rangle_{\psi(\xi, \kappa)}
  &=& \langle {\mathbf k}^\mu
  \rangle_{ref} +
                    \Delta k^\mu
  \end{eqnarray}}
  \item{spreadless transport of states:
  \begin{eqnarray}  
  \delta {x}_{\psi(\xi, \kappa)}^\mu  
 = \delta {x}^\mu_{ref} = const. \cr \cr
  \delta {k}_{\psi(\xi, \kappa)}^\mu 
 = \delta {k}^\mu_{ref} = const. 
  \end{eqnarray}}  
 \item{$x$-$k$ evolution~\cite{ref:ehrenfest} equations:
  \begin{eqnarray}
  &\, & \frac{d\langle {\mathbf x}^\mu \rangle}{\| d \langle
 {\mathbf x} \rangle \|}
  = \frac{\langle {\mathbf k}^\mu \rangle}{k_C}
   \cr \cr \cr
  &\, & d\langle {\mathbf x}^\mu \rangle \, d\langle {\mathbf k}_\mu
  \rangle = 0 \cr \cr
 &\, & \int_{path}
  \langle {\mathbf k}_\mu \rangle d\langle {\mathbf x}^\mu
 \rangle = extremal 
  \label{eq:kl-rr2}
  \end{eqnarray}}
  \item{path type constraints:
  \begin{eqnarray}
  \langle {\mathbf k}_\mu \rangle
  \langle {\mathbf k}^\mu \rangle &=& \pm k_C^2 \cr \cr
  \langle {\mathbf k}_\mu {\mathbf k}^\mu \rangle &=& \pm k_C^2 -
   \delta k_\mu \delta k^\mu
 \label{eq:kl-rr1}
 \end{eqnarray}}
 \item{{\em contact condition} between 
 the physically meaningful state-averages
 and the particle-trajectory ruling of the manifold:
  \begin{eqnarray}
  \langle {\mathbf x}^\mu \rangle_{\psi (\xi, \kappa) } =
  \xi^\mu \cr \cr
  \langle {\mathbf k}^\mu \rangle_{\psi (\xi, \kappa) } =
  \kappa^\mu 
  \end{eqnarray}}
 \end{itemize}
  Although no physical interpretation
 has been assumed so far for ${\mathbf k}$, it is evident
 that it corresponds to what is more traditionally known
 as
 4-momentum, ${\mathbf p}_\mu = \hbar {\mathbf k}_\mu$.

  Since geometric phase
 properties have been discussed mostly in the context
 of low energy phenomenae, the following
 will refer to 
 high energy aspects.
  Quantum states travelling  on  ``particle"-trajectories
  $\langle {\mathbf k}_\mu \rangle \langle
 {\mathbf k}^\mu \rangle  = const.$
  have two constants of motion:
  \begin{eqnarray}
    m_0^2 \, \buildrel \rm def \over = \, \frac{\hbar^2}{c^2}
    \, &\langle& {\mathbf k}_\mu \rangle \langle
    {\mathbf k}^\mu \rangle \cr \cr
    m_{bare}^2 \, \buildrel \rm def \over = \, \frac{\hbar^2}{c^2}
    \, &\langle& {\mathbf k}_\mu  
    {\mathbf k}^\mu \rangle 
  \end{eqnarray}
 the {\em rest} and {\em bare}
 mass of the state, related
 to each other by the spread
 of the state in $k$-space:
   \begin{equation}
    m_{bare}^2 -   m_0^2 =  \frac{\hbar^2}{c^2}
   \delta k_\mu \delta k^\mu
   \end{equation}
 a difference that for most stable systems is negative.
 The spread of $m^2_{bare}$ for an evolving
 quantum state is:
\begin{eqnarray}
 \langle \delta^2({\mathbf k}_\mu {\mathbf k}^\mu)
 \rangle_{\psi(\xi, \kappa)} =
 &\langle& \delta^2({\mathbf k}_\mu {\mathbf k}^\mu) \rangle_{ref}
  + \cr \cr
 &4& \, \| \Delta k \|
     \cdot \langle
     \delta(  {\mathbf k}_\mu {\mathbf k}^\mu) \delta(n_\mu {\mathbf k}^\mu)
     \rangle_{ref} + \cr \cr
 &4& \, \| \Delta k \|^2
     \cdot \langle \delta^2(n_\mu {\mathbf k}^\mu) \rangle_{ref}
 \label{eq:kl-def}
 \end{eqnarray}
 where 
  $\| \Delta k \|  \, \buildrel \rm def \over =\,
  | \Delta k_\mu \Delta k^\mu
 |^{1/2}$ and  $n_\mu   = \Delta k_\mu /
 \| \Delta k \|$.
 Due to the minimum of the expression
 in the vecinity of
 $(\pm m_0c^2, 0)$ for sub-luminous and 
 $(0, \pm m_0 c)$ for supra-luminous trajectories, the 
 linear term in $\| \Delta k \|$
 vanishes and the
 Klein-Gordon equation holds with good approximation:
 \begin{equation}
 {\mathbf k}_\mu {\mathbf k}^\mu \simeq const. \cdot
 {\mathbf 1}
 \end{equation}
 For high boost factors $\gamma \rightarrow \infty$ 
 however, the spread in $m_{bare}^2$
 diverges even if $\Delta m_{bare}^{ref}$ = 0, the Klein
 Gordon equation loosing accuracy:
   \begin{equation}
    \frac{\Delta m_{bare}}{m_{bare}} \simeq 
   \frac{\hbar \sqrt{2\gamma}}{m_{bare}c}
   \sqrt{\langle \delta^2({\mathbf k}_0 - {\mathbf k}_\parallel)
  \rangle_{min}}
 \label{eq:kl-eff}
   \end{equation}
 as the state approaches the light-cone and overlaps
 with the densely bunched $m_{bare}^2$ paths in this region
 of $k$-space, as well as with the supra-luminous states
 across the light-cone. This
 should be
 distinguished
 from seeing the state from a different system of reference
 (Lorentz boost).  The $\Delta m_{bare} / m_{bare}$ magnitude of the effect
 is on the order
 of 0.2\% 
 for a 1 eV/c wide $e^-$ state accelerated to LEP2
 energies, respectively 4\% for a 1 MeV/c wide $p$
 state accelerated to TEVatron energies.
 At E $\simeq$ 300 GeV
 a generic 1 eV/c wide $e^-$ state 
 overlaps with 
 hypothetical supra-luminous~\cite{ref:supra-luminous}
 components of $m_{bare}$ as high as 0.7 MeV/c$^2$.

   In summary, $dx_\mu
  dk^\mu = 0$ integrable trajectories
 have been shown to transport
 quantum states non-dispersively in $(x, k)$ projective space.
 The geometrical phase
 associated with the trajectories is extremal,
 its expression being that of the
 Classical Mechanics
 action.
  The trajectories
 are described by a constant of motion $k_\mu
 k^\mu = k_C^2$, more traditionally
 known as the ``rest mass",
 $m_0^2$.
  Highly boosted  quantum states 
 overlap both with
 higher $m_{bare}^2$ as well as with negative $m_{bare}^2$ states.

  I am thankful for the hospitality during completion of this
 work to the High Energy Physics group
 of the Wuppertal University - 
 under an Alexander von Humboldt Foundation grant, and to
 the Physics Department of the University of Colorado at Boulder.


\begin{references}
\bibitem{ref:dima} To be published. In essence it is possible
 to arrive at the ${\mathbf x}_\mu$
 and  ${\mathbf k}_\mu$ operators and their commutation relation
 solely on grounds related to separability of states
 and Differential Geometry properties of manifolds,
 without prior knowledge
 of physical equations.
\bibitem{ref:supra-luminous} Hypothetical supra-luminous
 transformations  connect  transformations across the
 light-cone, changing the sign of the pseudo-norm. Such
 type of action interchanges temporal 
 with spatial information parallel to the direction of boost,
 while rendering arbitrary information perpendicular to it.
   Supra-luminous transformations
   would hence obey  
 $\Lambda^\dagger G  \Lambda = - C_\parallel G $
 and have the form:
 \begin{equation}
 \Lambda^{ \cdot \mu}_\nu =
 \gamma' \left( \begin{array}{cc}
 1 \, \, \, \, \,  &    \vec{\beta}           \\
  \vec{\beta}\, \, \, \, \,  &  {\mathbf C}_\parallel
 \end{array} \right)
 \end{equation}
 where $\gamma' = 1/\sqrt{\beta^2 - 1}$ and
 $C_\parallel$
  a tensor selecting the parallel component
 to the boost.
\end{references}
 \end{document}